# Human-Centered AI Product Prototyping with No-Code AutoML: Conceptual Framework, Potentials and Limitations

**Mario Truss (mtruss@adobe.com)[1], Marc Schmitt (marc.schmitt@siemens.com)[2]**
[1]Adobe, [2]Siemens

**ABSTRACT**

This paper addresses the complexities inherent in AI product prototyping, focusing on the challenges posed by the probabilistic nature of AI behavior and the limited accessibility of prototyping tools to non-experts. A Design Science Research (DSR) approach is presented which culminates in a conceptual framework aimed at improving the AI prototyping process. Through a comprehensive literature review, key challenges were identified and no-code AutoML was analyzed as a solution. The framework describes the seamless incorporation of non-expert input and evaluation during prototyping, leveraging the potential of no-code AutoML to enhance accessibility and interpretability. A hybrid approach of combining naturalistic (case study) and artificial evaluation methods (criteria-based analysis) validated the utility of our approach, highlighting its efficacy in supporting AI non-experts and streamlining decision-making and its limitations. Implications for academia and industry, emphasizing the strategic integration of no-code AutoML to enhance AI product development processes, mitigate risks, and foster innovation, are discussed.

## 1 INTRODUCTION

In today's rapidly evolving technological landscape, the development of artificial intelligence (AI) products, such as classification models, has become integral to the success of organizations across various industries. Therefore, AI has become a competitive factor and a major area of investment as companies see the potential to improve their efficiency. However, the path to creating effective and human-centered AI products is often fraught with challenges related to knowledge gaps, uncertainty about product value, user acceptance, model development, time-to-market, security, reliability and more, which can cause most AI products to fail [1], [2], [3], [4], [5].

Prototyping has emerged as a viable method for validating the related risk to AI products before productive development resources are invested. However, most AI prototyping approaches allow for ideation of AI product idea by gaining insights into requirements for functionality or the desired experience of AI and to validate the user experience (usability) and marketability (desirability). Though, only limited conclusions about the performance, feasibility, usability and viability of the AI product in its real context can be made with existing approaches [6], [7], [8], [9]. This is the case because most prototyping techniques validate the product idea in theory or a simulated version of the final product, e.g., in the form of a clickable prototype. This is especially problematic, because the full functionality of AI products is dependent on the algorithm choice and data. Therefore, the feasibility of an AI product is not only limited to the capabilities of the AI engineers and needs to be differentiated from classical software products, which are as good as the code used to write them. The combination of the two factor is necessary to realize a functional AI product. This makes the outcome of such developments less predictable and therefore creates a greater risk for development organizations [10], [11].

Due to the inherent risk of AI products, e.g., created by malfunction and misinvestment, the demand for prototyping technique is high in these areas and solutions are sparse. This paper explores a cutting-edge approach to AI product prototyping by deploying the capabilities of no-code AutoML (NC AutoML) to support the AI product prototyping process, because NC AutoML has been sparsely discussed in prototyping research, which proves a research gap and justifies the need for a solution for AI prototyping. The potential to improve the AI product prototyping process by democratizing the core of AI product prototyping, the AI model development, for individuals with diverse background (AI experts and AI non-experts), will be analysed and documented in a conceptual framework (artifact). We will address the following research questions:

- *RQ1*: Can NC AutoML be integrated into the prototyping process of human-centered AI products?
- *RQ2*: What are potential of integrating NC AutoML in the AI product prototyping process?
- *RQ3*: What are limitations of using NC AutoML in the AI product prototyping process?

The goal is to uncover the considerable advantages, challenges, and limitations that organizations may encounter when adopting NC AutoML for AI product prototyping. To structure the method and paper, the design science research approach in the interpretation by [12] was used with the objective of creating a validated framework to showcase how NC AutoML can be integrated in the AI product prototyping process. The paper is structured as follows: Section 2 provides a comprehensive review of the literature, highlighting the significance of AI product prototyping, explaining NC AutoML, and presenting approaches to and challenges in AI product prototyping. Section 3 outlines the research methodology employed (DSR). In section 4, the framework development process is presented. Section 5 describes the potential, challenges, and limitations of using NC AutoML for AI product prototyping by examining a real-world case study and an evaluation. In section 6, the future research is discussed and in section 7 we conclude the research findings.

In a landscape where innovation and speed-to-market are paramount, the integration of NC AutoML into AI product prototyping processes can be transformative. This research aims to shed light on the opportunities and complexities associated with this emerging paradigm, offering valuable insights for researchers, practitioners, and decision-makers navigating the AI-driven future.

## 2 BACKGROUND AND LITERATURE REVIEW

### 2.1 Literature Review Process

To motivate and justify the framework development and research, first the relevant background theories were identified and then problems in AI prototyping for AI non-experts were derived from the literature by having used a literature search and a literature review with a qualitative content analysis approach, inspired by [13] [14] and [15]. All relevant literature was identified via Google Scholar, Science Direct, Emerald, ACM Digital Library, IEEE Explore and Semantic Scholar with the following keywords: "no-code" AND ("artificial intelligence" OR "AI" or "machine learning" OR "ML") AND ("prototyping" OR "prototype"). Additionally, backward, and forward searches and the tools Connected Papers and Researchrabbit.ai were used to identify additional literature. The literature search resulted in 1590 publications, which were subsequently filtered down via inclusion criteria (discusses AI product prototyping) and by reading the title and the abstract, which reduced the amount to 644. By scanning the articles, ultimately 48 relevant documents (research articles and book chapters) were identified that provide concrete insights into the challenges, approaches and potentials of AI prototyping with a focus on no-code approaches. The literature was then analysed inductively to extract theoretical theories, approaches and challenges in AI product prototyping for AI non-experts, which are presented in the section 2.2-2.5.

## 2.2  Human-Centered AI Product Management and Prototyping

AI product management is performed for products that are mainly enabled through data science and artificial intelligence (AI) [8]. These products are developed by pure AI teams or interdisciplinary teams consisting of AI experts, developers, designers, user experience researchers and a product manager. The product manager is responsible for the vision and strategy around the product and the development of the product, making this position highly relevant for the commercial success of products. This perspective enforces the realization of value. This value must be so high that customers are willing to pay more for the product than for the product provider to bring it to the market. If the product can be proven to be valuable, an investment decision toward developing the AI product will be made – or vice versa [16], [17], [18]. A common practice for evaluating value and market demand of product ideas is prototyping, as shown in Fig. 1.

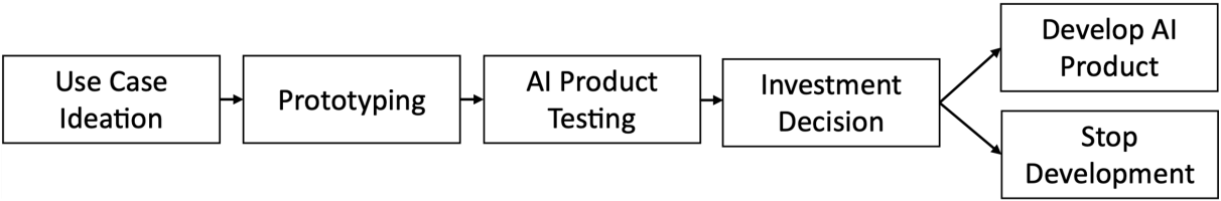

**Fig. 1. AI Prototyping Phase (Own Fig.)**

A prototype is described as a vehicle for going from a product idea to a market-fitted product by validating the chances of surviving against competition [19] and ensuring digital responsibility [20], [21]. According to Floyd 1984, prototypes serve to "enhance the communication between developers and users concerning the suitability" of the interaction "between system functions and work tasks". In a way, prototypes act as "boundary objects" to create a shared understanding of a problem space, which is especially necessary in the "phase of requirements elicitation" to reduce the "black box" characteristics of IS development processes [22]. This approach seems to be necessary for demonstrating the efficiency of the system from a business, technical and user perspective. The perquisite to having an adequate prototype is that it should be "work in such a way that it can be demonstrated to the users" given the "available resources" [6]. Research from University St. Gallen showed that this has become increasingly important for AI products, because „there is still not enough knowledge about artificial intelligence in companies" [7].

Prototypes usually come in the form of visual prototypes to evaluate user acceptance and/or functional prototypes, which allow customers to evaluate the product and its functionality to inform the customers' purchase decision-making process. Prototypes can be used to evaluate whether a product idea fulfils these product success criteria [7], [9], [19]:

- Viability (business view): Will this product be economically successful?
- Feasibility (technical view): Can this product be technically realized /built?
- Usability (user view): Can this product be used by potential users?
- Desirability (market view): Will there be paying customers?

The experimentation-heavy character of prototyping is deemed to be critical for ensuring that the AI product developed is human-centered (HCAI), which entails aspects of user orientation, ethics, safety, trustworthiness and responsibility [5], [8], [23], [24], [25], [26], [27], [28], [29]. Specifically, feasibility testing and reliability are more difficult to apply to AI products, because AI products are not directly programmed, but instead trained with historical data to learn how to solve problems [11]. This shows that considering only common errors in non-AI products, such code errors (incorrect code, e.g. wrong or missing logic) and user errors (incorrect usage, e.g. wrong input), are not enough to ensure feasibility for AI products [20, p. 8]. Factors such as data quality and AI-human interactions influence the functionality and reliability of the system and therefore also the human-centricity [4], [26], [27], [30]. Due to this, the prediction of product value and the customer willingness to pay for the AI product is hindered [20]. All these problems are reasons why leading academic institutes and big tech companies, prioritize research on the development of human-centered AI products [31], [32], [33], [34], [35], [36], [37], [38].

## 2.3 No-Code AutoML for Human-Centered AI Products

The traditional AI development process, more specifically machine learning development, is complex and requires extensive AI knowledge, including coding and AI knowledge. All ML development processes start with data collection (collect data) and are then followed by data preparation (clean and transformed data in necessary format). A ML model is only as good as the data used to train it; therefore, data quality poses a high risk for the development of dysfunctional or unethical AI solutions. Data collection is considered to be the most tedious of all the steps in ML model development [7], [39]. Then the ML algorithms need to be selected and the model training, validation, optimization, and evaluation need to be done by an AI expert, because ML expertise and coding language such as R or Python are required. Same applies for deployment.

No-code AutoML (NC AutoMM) is an approach, shown in fig. 2, for allowing AI non-experts without coding competency to develop machine learning models that drive AI products by providing a graphical user interface (GUI) and system guidance [40]. The main reason for this is the democratization and accessibility improvement of the AI product development process, because there is currently a gap between AI expert developers and AI non-experts [41]. The no-code idea is highly liked to the idea of citizen development and digital divide mitigation, which aim to equip people without coding knowledge to develop applications [21]. Past research has shown that AutoML is deemed to be significant productivity enhancer[42]. Furthermore, AutoML-like solutions are also mentioned as a tools for the design of human-centered AI products, due to its training automation, evaluation and testing capabilities and the accessibility improvement of AI product development for AI non-experts. [43], [44].

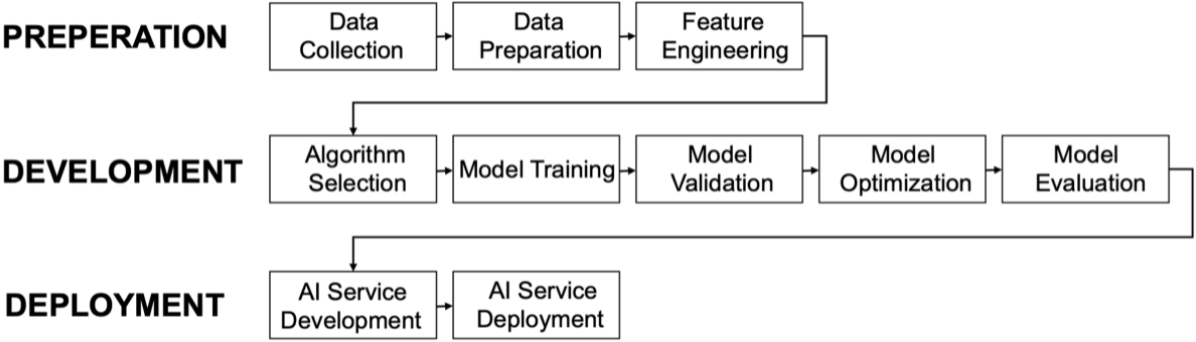

**Fig. 2. AutoML Process [40], [45], [46]**

„AI Platforms are the middle ground between an off-the-shelf-package and a bespoke build. They are provided by the large tech companies such as IBM, Google, Microsoft and Amazon, some large outsourcing providers, such as Infosys and Wipro; and specific platform vendors such as H20, Dataiku and RapidMiner" [47]. Established in 2012, DataRobot was among the early leaders in the field of NC AutoML. Subsequently, other enterprises like H2O, which introduced their Driverless AI platform, and Google, which offers technologies like Cloud AutoML, entered the scene in 2017 and 2018. This trend has expanded to include various cloud providers, such as Azure Machine Learning Studio AutoML and the AWS SageMaker Autopilot [48].

NC AutoML solutions need to be differentiated from code or low code AutoML solutions such as AutoGluan and AutoKeras, as well as from no-code AI as a service tools and platforms, which offer domain-fixed functionalities, such churn prediction or knowledge extraction (e.g. Adobe Sensei, ikigailabs.io, Gyana, or enhencer) [45], [49].

## 2.4 Challenges in AI Product Prototyping

The following challenges were inductively derived from via literature analysis, described in 2.1.

*C1. Missing knowledge AI capabilities and limitations*

The greatest challenges in AI prototyping are due to knowledge imbalances between stakeholders in the AI development process, mainly AI experts and AI non-experts, such as designers or business stakeholders [11], [50], [51], [52], [53], [54], [55], [56], [57].

*C2. Lack of boundary objects for effective collaboration*

This problem seems to be intensified by ineffective communication between these parties, which is mainly attributed to the lack of a *boundary object*, i.e., a representation or abstraction of the AI product idea, that could bridge the knowledge gap between AI experts and AI non-experts [7], [8], [11], [49], [57], [58], [59], [60], [61], [62], [63], [64], [65], [66].

*C3. Unpredictable AI behaviour and lack of functionality preview*

Additionally, the probabilistic nature of AI, created by the dependency on the training data, seems to be an impediment as well and requires a functional ML model to preview the actual functionality [7], [11], [23], [56], [57], [57], [67].

*C4. Development complexity*

This seems to be intensified to by the complexity of the data preparation and model training. The ladder is further catalysed by the high dependency on AI experts, due to the lack of knowledge and lack of a way to integrate AI non-experts [7], [8], [10], [56], [63], [66].

## 2.5 Related Approaches for AI Product Prototyping

AI product prototyping is still a relatively new research area, but making it accessible to AI non-experts seems to be the top of mind for elite research institutions, for instance Google, IBM, Apple and University of St. Gallen, which signal the relevant of this topic for industry and research [7], [56], [58], [68], [69]. The first and most mentioned approach to AI product prototyping, mentioned in 2018 and the following years, was the use of the Wizard of Oz (WoZ) technique to prototype and test AI product ideas. WoZ allows a rule-based simulation that mimics the model's capabilities to reduce technical effort [8], [11], [52], [57], [70], [71], [72]. Other traditional prototyping methods, such as modelling, mockups, design thinking and interviews, were mentioned to inform AI product design [8], [73], [74]. Additionally, an AI playbook was mentioned as a tool for exploring common error scenarios of envisioned AI products by providing contextual, actionable guidance for simulating and testing those scenarios [75].

A more cutting-edge approach to AI prototyping is the use of pre-trained ML models. For instance, the use of pre-trained object recognition AI to embed it in an app has been discussed multiple times [59, p. 1084], [72], [76], [77]. A special form of this approach was the integration of large language models (LLMs), e.g. to use of generative AI to explore and ideate products and then generate HTML and CSS code for high fidelity prototypes [78]. Three papers by Google Research further explored the use of LLMs to improve AI product prototyping through prompt-based prototyping, which allows prototyping an AI functionality with GPT [56]. Advancements to this were the combination of multiple LLMs to increase functionality, which is called "LLM chaining" [68] and the integration of the LLM responses into UI prototypes that to simulate the experience [69]. Similar prototyping approaches involving LLMs with wireframing of the WoZ method and the ability to integrate functional AI model have been used [55]. These approaches do not really solve the issue of creating a new AI for an unsolved problem because they built up on an existing AI model instead of creating a custom one, as in AutoML and traditional ML/AI platform.

Few articles have discussed the use of AI platforms, such as AutoML, to prototype AI products, but these are mainly written by developers and focused mainly on technical details, such as ML model training and performance metrics. Though they provide proof that AutoML can create comparable results to manual ML [39], [40], [42], [46], [79], [80], [81], [82]. Therefore, the findings are not centered around the needs of AI non-experts, but instead focus on improving the development process for AI experts, such as ML engineers [58], [66], [83]. Most of the research written with AI non-experts presented custom software solutions for AutoML-like approach or mentioned AutoML at the meta-level without providing evidence for its utility in the prototyping process [66], [84], [85], [86], [87], [88], [89]. One exception is one paper, which discuses a hybrid of AutoML-like technology and a pre-trained algorithm with the technology "Teachable Machines" by Google [65]. These insights show the potentials of using a publicly available AutoML solution as a tool for AI non-experts for AI product prototyping are yet to be discovered, therefore, a research gap was assumed.

## 3 MATERIALS AND METHODS

The goal of this research was to determine the answers to RQ1, RQ2 and RQ3. To ensure the scientific validity of results, the, in Information Systems Research (ISR) highly established, Design Science Research (DSR) paradigm was used as a research approach, as shown in Fig. 3.

The goal of DSR is to produce descriptive knowledge in the form of an *artifact* that is relevant to the domain of the problem discussed in the related research, in our case AI product prototyping [90]. For this research, a *framework* was selected as a desired artifact because it can showcase the utility of artifacts at the meta-level [91], [92].

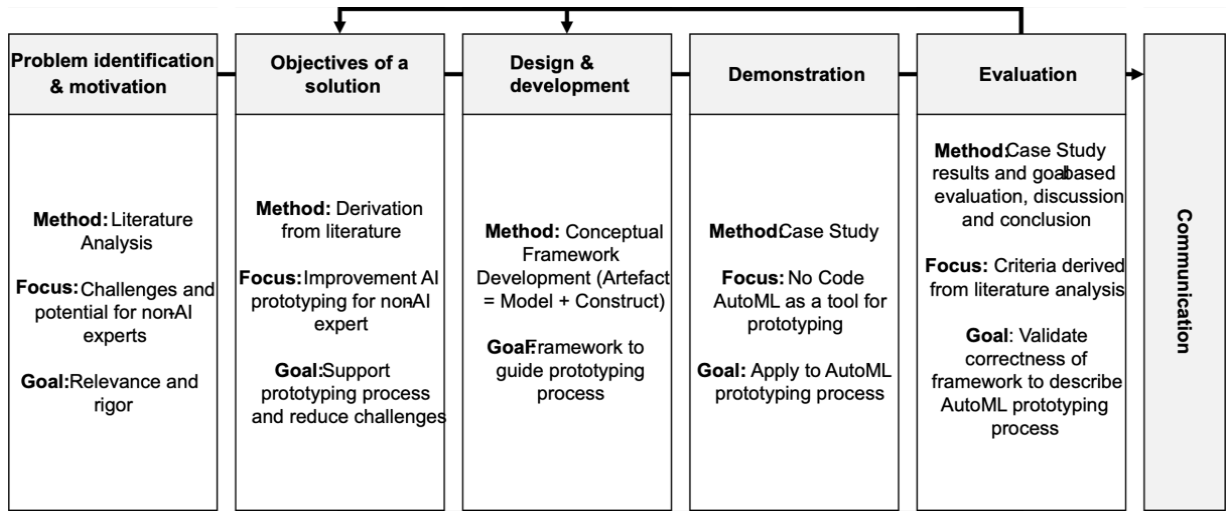

**Fig. 3. Overview of DSR approach [90]**

To evaluate the artifact created through the DSR process, the solution for the research problem must be an improvement [90]. The challenges described in section 2.5 and the unmet need for a solution, described in section 2.6, were used to justify the development of a solution for AI product prototyping for AI non-experts with AutoML, which aims to allow access to the ML development process for AI non-experts. Based on these insights, the solution objective of this research was to develop a framework that explains how AutoML can improve the AI product prototyping process, especially for AI non-experts.

To demonstrate the utility of the framework and the impact on its context, a case study, which is a recommended method in DSR, was used for demonstration and evaluation [91], [92], [93]. The case study was conducted with Google AutoML, as it had one of the best performances in previous attempts to train productive models and because it has not been mentioned by AI non-experts as a means for prototyping [81], [94], [95], [96]. This approach was crucial for ensuring the generation of new knowledge. Additionally, product success criteria were derived from prototyping theory (see 2.1), and the challenges identified in the literature review (see 2.3 to 2.3; C1 to C4) were used to evaluate the case study results and the framework for its utility in supporting the AI product prototyping process as follows:

- *A. AutoML integrates into the AI product prototyping stages (2.2):*
    - o A1. Ideation
    - o A2. Prototyping
    - o A3. Testing
- *B. AutoML provides additional insight into the product success criteria (see 2.2):*
    - o B1. Viability
    - o B2. Feasibility
    - o B3. Usability

- o B4. Desirability
- *C. AutoML solved or improved these challenges (2.4):*
  - o C1. Missing knowledge AI capabilities and limitations
  - o C2. Lack of boundary objects for effective collaboration
  - o C3. Unpredictable AI behaviour and lack of *functionality* preview
  - o C4. Development complexity

If these criteria are being met, the solution is considered to have *validity* (artifact works), *utility* (provides value to more than the prototyping process), *quality*, and *efficacy* [12]. This hybrid approach of combining naturalistic (case study) and artificial evaluation methods (criteria-based analysis) ensures a rigorous evaluation of the proposed frame, which allows for a stronger repeatability, falsifiability, and internal validity, which in turn creates scientific reliability [90], [97], [98], [99], [100]. The DSR approach allows the structured integration of insights generated from research and from a case study to build a relevant and rigorous conceptual framework for AI product prototyping for AI non-experts (DSR artifact), as the main goal of DSR is to develop descriptive knowledge [90], [92], [101], [102].

## 4 RESULTS

### 4.1 Development Process

The insights from the literature review in section 2.5 showed that there is currently no solution to solve the challenges in AI product prototyping in section 2.4, as visualized in Fig. 4. Therefore, it is assumed that a solution for these challenges provides value, which justifies the development.

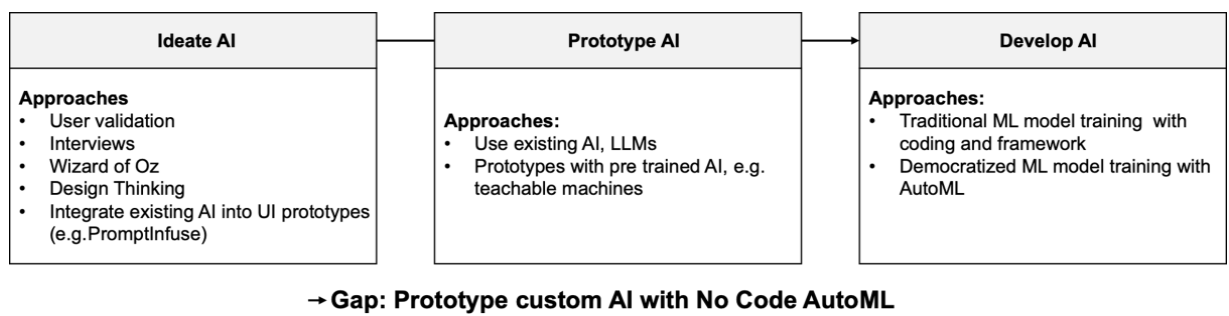

Fig. 4. Problem Identification with a Solution Gap for Artifact Justification

The final framework was developed in three iterations. The goal of the framework is to provide a process description of how NC AutoML can be integrated into the AI product prototyping process to improve the evaluation of the product success criteria and to reduce the challenges in AI product prototyping (see 3.). All decisions were based on the identified solution gap (see 3, 2.4 and 2.5). The design and development process of the framework was as follows:
- As a first iteration, the initial framework was created, which was solely a combination of the product success criteria and the AutoML process (see 2.2. and 2.3).

- Second, Then the challenges, that had been identified in the literature analysis, were integrated into the framework (2.4).
- Finally, through the case study, we added which aspects of a prototyping can be supported with AutoML. This resulted in the integration of the functionalities of AutoML technologies, which were derived from the case study in Google AutoML.

## 4.2  Final Framework

Based on the research results and design process, the conceptual framework, shown in Fig. 5 describes how AutoML can be integrated into the AI product process and how it can support AI non-experts with domain knowledge in the AI product prototyping. The final framework integrates the prototyping phase (see 2.2) with the product success criteria (see 2.2) into the AutoML process (see 2.3) to solve challenges of AI product prototyping (see 2.4) and fills the solution gap (see 2.5). The NC AutoML product prototyping process, described in the framework, starts with ideation, and ends with the AI product investment decision, which determines whether an AI product has enough value to be evaluated. The framework aims to connect AI non-experts with an idea to AI experts, who develop the final solution by allowing them to participate in the AI product prototyping process.

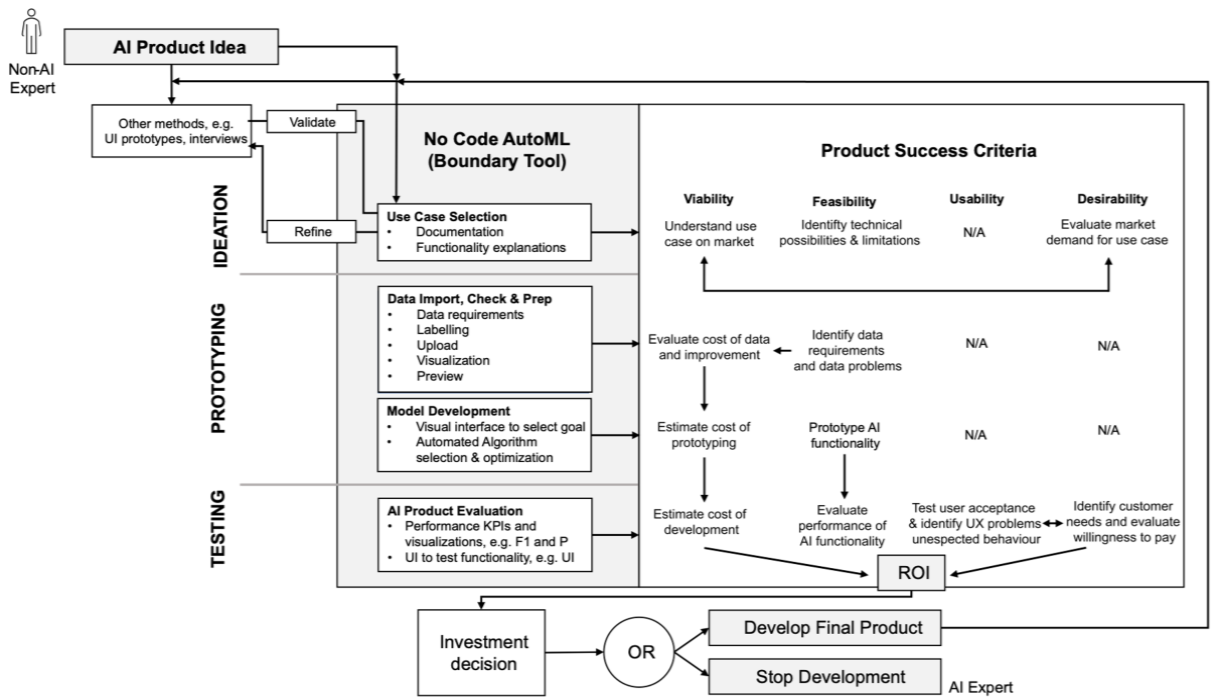

**Fig. 5. NC AutoML AI Product Prototyping Framework**

The framework (see Fig. 5) shows that NC AutoML is integrated into the AI product prototyping process as a boundary tool, which benefits the AI prototyping process in multiple ways:
- *Use Case Selection*: AI product ideas can be validated based on the use case by matching the product idea with the AI capabilities that ML is able to support. This can be realized by using the explanations in the tool or using the documentation to *validate* and *refine* existing

- prototyping methods, e.g., UI prototypes, as it allows transparency about the capabilities and limitations (feasibility). Additionally, the insights generated can be used to compare the capabilities with the market offerings to prove a gap (feasibility) and match it with the customer demand and functional needs (desirability). This approach is aimed to support the AI product *ideation*.

- *Data Import, Check & Prep*: The otherwise code-intensive process of importing data, validating, and pre-processing it for model training is being made accessible to users without coding capabilities by allowing a simple *file upload* and showing guidelines on file *requirements*, e.g., format or data volume (feasibility). Through *visualizing* and *previewing data*, *data problems* can be quickly identified and corrected by *labelling* (feasibility). Based on this, the cost of data acquisition, for missing data and the improvement of existing data can be evaluated (viability), which is needed to prove that the data is usable (usability and desirability). This is aimed to support the AI product *prototyping*.

- *Model Development*: Model training, a process which requires both coding and AI expertise and is deeply rooted in math and computer science, can be made accessible by *automating* all necessary areas, e.g., *algorithm selection and optimization*. This is realized with an accessible *visual interface* for AI non-experts. Through this approach, the cost of prototyping is reduced by lower entrance requirements (viability) and the general possibility of prototyping AI functionality, which allows faster testing results (feasibility). This is aimed to support the AI product *prototyping*.

- *AI Product Evaluation*: The core of the AI product prototype, the ML model, can be evaluated because of the steps mentioned before. Instead of relying on a fictitious AI functional, which might be misleading, the real ML model proves to a certain extent, what is currently possible with the data and state of the art AI methods. With *performance KPIs*, e.g., precision or accuracy, and *visualizations*, e.g., confusion matrix, the performance and functionality of the AI product can be evaluated (feasibility). Through a *user interface* (UI), the interactions with AI can be tested with user by allowing users to observe the output generated by the AI. This approach can provide insights into the user acceptance of the AI product functionality, potential problems, and unexpected behaviour (usability). These insights can then be used to evaluate the match between the produced AI functionality and the *customer needs* and to evaluate the *willingness to pay for the functionality* (desirability). Based on these insights into the current limitations of the AI, the estimated cost of development can be derived (viability). This approach is aimed to support the AI product *testing*.

These insights generated with NC AutoML can then allow return on investment assumptions, which can be used to influence the investment decision (develop final product vs. stop development). Insights from AI product testing are also assumed to be beneficial to improve the prototype iteration, e.g., potential user does not accept the solution because there is bias in the data, which would lead to an improve of the data and a new model, which then could be tested again.

## 4.3 Evaluation with Case Study of No-code AutoML

To validate the ability of no-code AutoML (NC AutoML) to support the AI product prototyping process, Google Vertex AI, now referred to as NC AutoML. The NC AutoML user had no prior knowledge of ML and no coding capabilities for machine learning model development, therefore, they totally relied on system support by NC AutoML. The AI product idea, used as a case study example, is the real-time classification of customer requests, to identify tickets, that need to be directed to subject matter experts (escalation). We used a dataset with 3440 training examples. In the UI of NC AutoML, the following AI use cases were presented and could be selected and further researched with the provided documentation [103]:

- Classification, entity extraction, and sentiment analysis (emotion analysis) of text and ticket categorization
- Regression, and classification of tabular data, e.g.
- Classification, detection, action detection, classification, and object tracking of images and videos.
- Generative AI for text, chat, code, and image generation

Uploading was seamlessly possible in files formats, such as CSV, BigQuery table (data warehouse, JPEG, PNG, TXT, MOV, MPEG-4, MP4 and AVI. For the selected use case, the CSV upload was used. Afterwards, the data was previewed, the class balance was checked, and the minimum requirements were reviewed from the documentation. To inform the user, the system offers information on best practices for AI development and suggestions on how to improve the dataset, with additional links to the documentation. An incorrectly identified label could be corrected with the internal data labelling functionality.

Model training, so the main part of prototyping of the AI, was automatically performed by the system. The only decision that the human must make is to decide how much of the data should be used for training, normally 80%, and how much for model validation and evaluation, the rest of the data. Each decision was supported with UI explanations and links to the documentation. For the dataset of 3443 examples, AutoML required 5 hours to train the ML model which enabled the AI functionality.

The results of the ML model are presented in a KPI overviews, which is shown in Fig. 6. The metric used for the text classification use case described were average precision, precision, recall, F1 score, area under curve (AUC) and receiver operating characteristic curve (ROC), which are further explained in the UI and state of the art in ML development. Additionally, three solutions were offered to understand the predictions. First, there is an option to adjust the confidence threshold of the prediction, which allows user to understand the relationship between the predictive KPI and the confidence. Second, the feature importance, e.g., the description text that drives the classification decision of AI, is shown to the user. Third, a confusion matrix can be utilized to understand which categories are harder to predict for the AI. Additionally, the confusion matrix insight was supplemented by examples for when a prediction was right and when it was wrong [103].

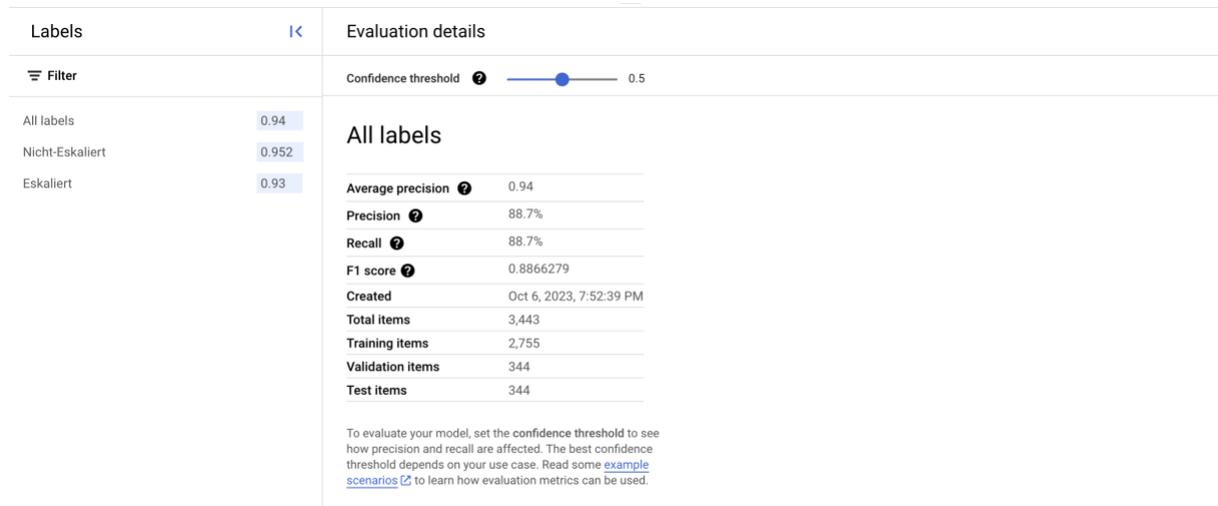

Fig. 6. Example of Google AutoML Evaluation View

The trained model could then be exported in the TensorFlow format or tested in the UI. For this purpose, the model needed to be deployed and could then be accessed by a UI that allowed the manual input of data – in this case a customer support query – as visible in Fig. 7, and showed the output, e.g., classification of ticket by need escalation.

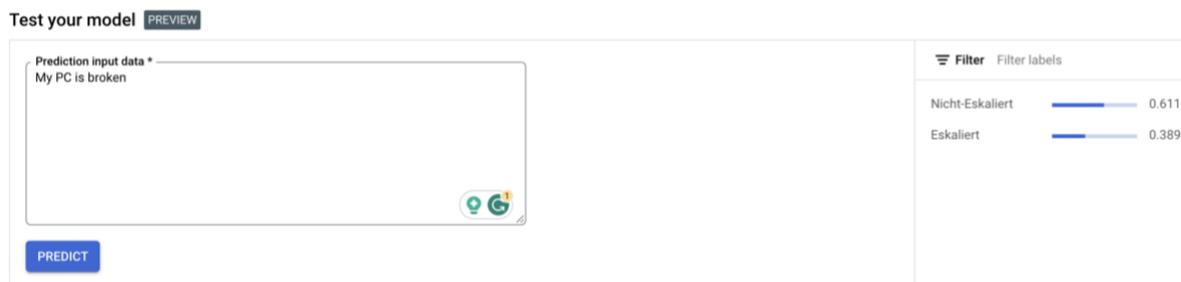

Fig. 7. Google AutoML Preview

The interactivity of this preview UI allowed the AI non-experts to interact with the AI to understand its predictive capabilities and limitations. For each input the UI showed the confidence for the prediction, which allows to understand which prediction are harder to achieve for the AI, compared to other queries. Additionally, the generated prediction can be accessed via an API endpoint. To allow generalizability, it was further validated that this type of interactive AI product testing UI is also provided for regression, image, and sound classification in Google AutoML. Further, it was validated that comparable commercial solutions, such as AWS Sagemaker Canvas and Azure AutoML offer comparable functionalities [104], [105].

## 4.4 Criteria-based Evaluation

Based on the identified product success criteria that are evaluated with prototyping techniques and the identified challenges in AI product prototyping, the utility of NC AutoML and the presented functions in the conceptual framework will be analysed in the following section (details about evaluations methodology in section 3).

*A1-A3. Integration in Prototyping Stages*

The case study showed that NC AutoML integrates into the AI product prototyping stages (2.2). Ideation was mainly supported by documentation and UI explanations in AutoML. Prototyping is supported by making all manual steps accessible through a UI and automating all tasks that require AI knowledge, e.g., algorithm selection and model optimization. The preview function additionally made it possible to interact with and test the AI to understand its capabilities and limitations.

*B1. Product Success Criteria: Viability*

The market and demand checks are assumed to generate valuable insight for validating the viability of the product idea. This is made possible through the potential costs that might arise, e.g., through improvement of the data, and the expected development effort. These insights can be generated through the requirements, performance metrics, errors and biases that can be discovered in the prototyping process. Examples include the discovery of incorrect classifications in the confusion matrix or the use of the NC AutoML UI to interact with the AI. Furthermore, the insights into the other product success criteria insight, e.g., desirability via customer feedback on the value provided and willingness to buy, based on interactions with prototypes, can further inform the commercial evaluations to inform ROI predictability. Additionally, it can be assumed that the feasibility can be further evaluated by proving a market gap through comparing the developed ML model with existing solutions in the market.

*B2. Product Success Criteria: Feasibility*

The Given the fact, that NC AutoML delivers comparable results to traditional ML implementation (see section 2.3), it can be assumed, that the feasibility of the AI product idea can be properly validated based on the given data. This allows AI non-experts to quickly validate whether their AI product idea can be solved with state-of-the-art AI approaches and the current resources available, in specific with the current data available. Through this approach, AI risks such as bias and algorithmic errors can be identified early in the development phase. This can further inform whether additional resources are needed to improve data quality, e.g., via data engineers, or custom AI algorithms need to be developed, which requires access to AI experts.

*B3. Product Success Criteria: Usability*

Insights about usability can only be directly gathered for AI non-experts through the AI product testing capabilities, e.g., the preview UI, where a user can interact with the AI. Through the API endpoint offered by NC AMT, the generated prediction can be further integrated into a different user interface to evaluate whether the generated output can be integrated into the final design of the indented product. The ladder is mostly not accessible as person without coding skills. Nevertheless, both methods allow the early discovery of user acceptance, usage problems and unexpected behaviour of the UI to inform future improvement and development of the ML model, that drives the AI.

*B4. Product Success Criteria: Desirability*

The desirability of an AI product idea can be evaluated by using the prototype as a tool to evaluate the match between the current capabilities and the customer needs. Though it must me mentioned that the preview UI to preview the AI functionality might not be the most suitable user interface to the validate user desirability. Additionally, the willingness to pay could be evaluated by integrating suitable interview questions in the AI product prototyping test, e.g., to evaluate perceived quality and experience of the AI functionality, e.g., prediction, enabled by the ML model. With this interface, an interviewer could then ask how valuable the functionality is for that user.

*C1. Challenge 1: Missing knowledge AI capabilities and limitations*

The case study showed that NC AutoML provides UI guidance and additional documentation on best practices, supported use cases, capabilities, limitations, and guidelines for the development of ML models – the core of most AI products. The UI of NC AutoML bridges the knowledge gap of AI non-experts, as all needed knowledge is provided visually in real-time to the AI non-experts and the development can be fully done via the UI. Additionally, the testing functionality of NC AutoML, e.g., the preview UI, further allows for the understanding and evaluation of the AI capabilities and limitations. The confusion matrix and similar tools additionally provide transparency on which cases AI can provide a correct prediction and for which there are problems in the prediction. Therefore, it can be assumed that NC AutoML provides significant solutions for challenge 1.

*C2. Challenge 2: Lack of boundary objects for effective collaboration*

As the complete ML model development process is accessible without code and all AI-related parameters are explained in the UI, it can be assumed that NC AutoML acts as a boundary object and therefore makes the collaboration between AI experts and AI non-experts more effective. Therefore, AI product prototyping has become increasingly accessible and democratized, bridging the gap between ideation and the final development of the AI products.

*C3. Challenge 3: Unpredictable AI behaviour and lack of functionality preview*

Because NC AutoML uses real data, it can be assumed that it will create transparency about the real functionality of the AI given the current resources, including the capabilities and limitations. This is enabled through the evaluation capabilities, such as performance metrics and the possibility of interacting with the AI through a UI. This can can help mitigate security and reliability threads, such as biased data or under- and overfitting, as well as error due to bad model quality and lack of data. Though, it must be noted that the NC AutoML UI might not represent the exact future AI product experience, due to not being fully integrated into the final UI.

*C4. Challenge 4: Development complexity*

It can be assumed that NC AutoML reduces parts of the complexity of the development process, because it reduces the requirements for participation in the development by providing a no-code interface. Nevertheless, ML model development is a complex task that requires the learning of the AI-related topics, e.g., the meaning of evaluation KPIs. Additionally, it needs to be mentioned that NC AutoML created transparency about potential data problems, but only simple problems such as labelling can be solved directly in NC AutoML. More complex data problems, such as volume, bias or syntax problems in the data need to be solved outside of NC AutoML with data collection and engineering methods.

*Validity, Utility, Quality, and Efficacy of the Framework*

Given that NC AutoML can be integrated in the prototyping stages (A1-A3), allows insights into all four product success criteria (B1-B4), and provides solutions to the challenges (C1-C4), it can be assumed that NC AutoML can be integrated into the AI product prototyping process with positive effects. Therefore, the framework created (DSR artifact) is assumed to fulfil the criteria of validity, utility, quality, and efficacy.

# 5 DISCUSSION

## 5.1 NC AutoML for Human-Centered AI Product Prototyping

Given the generated insights, NC AutoML has been proven to be a valuable tool in the AI product prototyping process for evaluating product success criteria, such as viability, feasibility, desirability, and usability, which are needed to reduce risk and allow estimations of the potential ROI of AI products. It can support all three stage of the human-centered AI product prototyping process (ideation, prototyping and testing). (RQ1, RQ2).

As the currently available methods can only provide limited evidence for these factors, using AutoML to quickly evaluate AI product ideas can help companies to make conclusions about the usability of the data for the ML model and therefore the feasibility of the AI product idea. Furthermore, the interaction with the AI product, the acceptance of the AI products and its ROI can be evaluated. This is an area in which traditional prototyping methods (see 2.4 and 2.5) have serious limitations. Therefore, this approach can help mitigate misinvestments in AI product ideas, which, for example are not able to be developed productively. Furthermore, the NC AutoML model can be used in future development to communicate business users' ideas and to get first benchmark data for future ML models. Additionally, there is a potential for NC AutoML to act as an interface between AI experts and AI non-experts, such as AI product users and business-oriented stakeholders to ensure human-centricity early in the AI product lifecycle, while reducing the (RQ2). These steps should be considered before staffing scarce development resources onto AI product ideas, which have been entirely validated, to ensure human-centricity.

Nevertheless, specific data-related task, e.g., data acquisition and improvement, as well as custom model development and the development of user interfaces that embody the AI functionality, which are needed for a complete AI product validation and full human-centricity, cannot necessarily be fully realized with NC AutoML. Also, the blackbox nature of ML model created with NC AutoML could be potential barrier to productionizing the AI product and to ensuring human-centricity, especially reliability, user acceptance and fairness, but also feasibility. This is aligned with other research, e.g. [106] Furthermore, it is not possible to test the holistic AI experience without coding knowledge, as the preview UI to test the ML model might have limitation with regards to the user interface and experience, compared to a customized UI, developed by UI/UX designers (RQ3).

## 5.2 Democratization of AI Product Prototyping

Democratized AI product prototyping through citizen development, particularly via no-code AutoML platforms, signifies a fundamental shift in the way technology is created and used. By empowering 'citizen developers'—individuals without extensive technical training—these platforms facilitate the creation of AI applications by a wider demographic, significantly diversifying the range of perspectives and needs addressed in AI solutions. However, while democratizing AI development can lead to more equitable technology access, it also necessitates robust ethical guidelines to ensure responsible use. Furthermore, the surge in citizen development potentially reshapes job landscapes, as traditional roles in AI development evolve to accommodate this new wave of innovators. Additionally, explainable AI becomes increasingly crucial in this context [25]. As non-experts create complex models, the ability to understand and trust AI decisions is essential for both developers and the end-users (RQ2).

## 5.3 Ethical and Social Considerations

Ethical and social risks, such as bias through data can be reduced through the product prototyping process, as documented in the framework. However, it must be noted that data-induced bias must also be evaluated in the data acquisition stages as well as in the productive AI product development, which is outside of the scope of NC AutoML (RQ3). Additionally, data privacy, accountability, consent and other human-centered AI topics should be considered before feeding them into ML models that drive AI product prototypes [107], [108], [109], [110]. Therefore, it must be considered that other AI product evaluation criteria that allow for the social and ethical impacts of products, e.g., the impact of an AI product on the employment of the citizens or the potential harm that the functionality could cause when being misused, must be considered.

## 5.4 Limitations of Research

Despite the scientific rigor of the utilized method, as justified in section 3, this research has the following limitations: The case study and the criteria-based evaluation were conducted only with:

1. The NC AutoML solution Google Vertex AI for the use case classification.
2. One dataset and one ML functionality type (classification) was used in the case study.

To improve the generalizability of these insights, other AI use cases and competitive NC AutoML solutions were cross validated. Nevertheless, the following steps are needed to allow full generalizability and should be done in future research:

1. Multiple AutoML providers should be fully evaluated to determine whether some NC AutoML solutions have a functionality gap that would reduce the support in the AI product prototyping process.
2. Multiple ML use cases should be evaluated.
3. User acceptance testing with AI non-experts, e.g., product managers, UX designers, etc., to evaluate the human factors in the human-centered AI product prototyping process.
4. Multiple AI use cases should be evaluated to evaluate whether NC AutoML provides identical benefits for all major AI use cases, that are driven by ML.

# 6 FUTURE RESEARCH

## 6.1 Human-Centered AI Product Development

Future research should explore the impact of NC AutoML on developing fully functional human-centered AI products. Building on the current understanding of in NC AutoML platforms, future studies should investigate how this democratization of AI impacts various sectors and demographics. There is a need to understand the long-term implications of enabling non-technical individuals to create AI solutions, particularly concerning job dynamics, skill requirements, and the evolving role of professional AI developers. Further research should also delve into the ethical dimensions of NC AutoML to examine how inclusive AI creation processes can be designed to uphold ethical standards and prevent misuse. Additionally, the role of explainable AI in enhancing trust and acceptance among NC AutoML users warrants detailed investigation. Understanding how these individuals interact with and interpret AI outputs is crucial for designing more intuitive and transparent AI systems. Overall, future research should aim to provide comprehensive insights into how citizen development within HCAI shapes the landscape of AI innovation, user acceptance, and the ethical use of AI technologies [8], [26], [27], [111].

## 6.2 Mitigating Code Vulnerabilities in AutoML Systems

As NC AutoML platforms become more accessible to non-technical users, the risk of unintentionally introducing vulnerabilities or misconfigurations increases. Research should aim to develop robust frameworks and tools that can automatically detect and address potential security flaws within AI models created by citizen developers. This includes exploring advanced algorithms that can identify vulnerabilities in real-time and suggest corrective measures. Additionally, there is a need to educate citizen developers, such as AI non-experts, about best practices in AI security, ensuring they are aware of potential risks and mitigation strategies. This research will not only enhance the security of AI systems but also reinforce user trust in AutoML technologies, aligning with the broader goals of Human-Centric AI development. By addressing these challenges, future advancements can ensure that AutoML systems are not only accessible and efficient but also secure and reliable [112].

## 7 CONCLUSION

In this paper, NC AutoML was evaluated as tool to improve the prototyping process of human-centered AI products. To prove the utility of NC AutoML in this context, a conceptual framework was developed and evaluated using the DSR method to show how NC AutoML can be integrated into the prototyping process of human-centered AI products to address current challenges for AI non-experts (see section 2.4), which have not been solved by current approaches (see section 2.5). The research shows that NC AutoML can be a valuable extension to commonly used prototyping methods to reduce the challenges currently discussed. Despite limitations (see section 5.1 and 5.4), this research provides significant insights on how, the human-centered technology, NC AutoML can improve the AI product prototyping process, reduce investment risks, and improve collaboration between AI expert and AI non-experts. NC AutoML automates the ML model training process, which is one of the core activities in AI product prototyping and makes it accessible to stakeholders without coding and AI expertise. The generated research insight provides a significant addition to the current body of knowledge by providing a framework for AI non-experts, which provides guidelines on the usage of publicly available NC AutoML technologies, which have not been discussed in the context of developing interactive AI product prototypes. This innovation is not confined to any specific sector; rather, it permeates all industries, empowering businesses, and organization. NC AutoML enables organizations to efficiently integrate AI into their processes, enhancing decision-making, optimizing operations, and fostering innovation. This inclusivity in AI development catalyses a new era of digital innovation and product, where the benefits of AI are more universally accessible and impactful across various domains. This can ultimately be a catalyst for interdisciplinary AI prototyping and development processes, driven by AI experts and AI non-experts, which improve the circumstances to develop human-centered AI products from an early stage.

**CRediT Author Contribution**

Mario Truss: Conceptualization, data curation, formal analysis, investigation, methodology, project administration, resources, software, supervision, validation, visualization, writing – original draft preparation, writing – review & editing. Marc Schmitt: Methodology, validation, writing – original draft preparation, writing – review & editing.

**Acknowledgments**

The authors would like to thank their employer for allowing them to do independent research. Thanks for Adobe Research for validating the research on scientific validity.

**Funder information**

No explicit funding.

**Declaration of interest statement**

The authors declare that they have no known competing financial interests or personal relationships that could have appeared to influence the work reported in this article. The paper was approved via an internal publication validation process to ensure scientific validity and independence of third-party copyright.

# REFERENCES

[1] P. Buxmann, T. Hess, and J. B. Thatcher, 'AI-Based Information Systems', *Bus. Inf. Syst. Eng.*, vol. 63, no. 1, pp. 1–4, Feb. 2021, doi: 10.1007/s12599-020-00675-8.

[2] McKinsey, 'The state of AI in 2022--and a half decade in review | McKinsey'. Accessed: Dec. 26, 2022. [Online]. Available: https://www.mckinsey.com/capabilities/quantumblack/our-insights/the-state-of-ai-in-2022-and-a-half-decade-in-review#talent

[3] McKinsey, 'The state of AI in 2023: Generative AI's breakout year | McKinsey'. Accessed: Nov. 02, 2023. [Online]. Available: https://www.mckinsey.com/capabilities/quantumblack/our-insights/the-state-of-ai-in-2023-generative-ais-breakout-year

[4] W. Liang *et al.*, 'Advances, challenges and opportunities in creating data for trustworthy AI', *Nat. Mach. Intell.*, vol. 4, no. 8, Art. no. 8, Aug. 2022, doi: 10.1038/s42256-022-00516-1.

[5] O. Ozmen Garibay *et al.*, 'Six Human-Centered Artificial Intelligence Grand Challenges', *Int. J. Human–Computer Interact.*, vol. 39, no. 3, pp. 391–437, Feb. 2023, doi: 10.1080/10447318.2022.2153320.

[6] C. Floyd, 'A Systematic Look at Prototyping', in *Approaches to Prototyping*, R. Budde, K. Kuhlenkamp, L. Mathiassen, and H. Züllighoven, Eds., Berlin, Heidelberg: Springer, 1984, pp. 1–18. doi: 10.1007/978-3-642-69796-8_1.

[7] W. Brenner, B. Van Giffen, and J. Koehler, 'Management of Artificial Intelligence: Feasibility, Desirability and Viability', in *Engineering the Transformation of the Enterprise*, S. Aier, P. Rohner, and J. Schelp, Eds., Cham: Springer International Publishing, 2021, pp. 15–36. doi: 10.1007/978-3-030-84655-8_2.

[8] A. Kore, 'Prototyping AI Products', in *Designing Human-Centric AI Experiences: Applied UX Design for Artificial Intelligence*, A. Kore, Ed., in Design Thinking. , Berkeley, CA: Apress, 2022, pp. 367–382. doi: 10.1007/978-1-4842-8088-1_8.

[9] T. Zwingmann, *AI-powered business intelligence: improving forecasts and decision making with machine learning*, 1st ed. Beijing Boston Farnham Sebastopol Tokyo: O'Reilly, 2022.

[10] Q. Yang, A. Scuito, J. Zimmerman, J. Forlizzi, and A. Steinfeld, 'Investigating How Experienced UX Designers Effectively Work with Machine Learning', in *Proceedings of the 2018 Designing Interactive Systems Conference*, Hong Kong China: ACM, Jun. 2018, pp. 585–596. doi: 10.1145/3196709.3196730.

[11] Q. Yang, A. Steinfeld, C. Rosé, and J. Zimmerman, 'Re-examining Whether, Why, and How Human-AI Interaction Is Uniquely Difficult to Design', in *Proceedings of the 2020 CHI Conference on Human Factors in Computing Systems*, Honolulu HI USA: ACM, Apr. 2020, pp. 1–13. doi: 10.1145/3313831.3376301.

[12] S. Gregor and A. R. Hevner, 'Positioning and Presenting Design Science Research for Maximum Impact', *MIS Q.*, vol. 37, no. 2, pp. 337–355, 2013.

[13] J. vom Brocke, A. Simons, B. Niehaves, K. Riemer, R. Plattfaut, and A. Cleven, 'Reconstructing the Giant: On the Importance of Rigour in Documenting the Literature Search Process', in *ECIS 2009 Proceedings*, Jun. 2009.

[14] S. Elo and H. Kyngäs, 'The qualitative content analysis process', *J. Adv. Nurs.*, vol. 62, no. 1, pp. 107–115, Apr. 2008, doi: 10.1111/j.1365-2648.2007.04569.x.

[15] H. M. Cooper, 'Organizing knowledge syntheses: A taxonomy of literature reviews', *Knowl. Soc.*, vol. 1, no. 1, pp. 104–126, Mar. 1988, doi: 10.1007/BF03177550.

[16] B. Namatherdhala, N. Mazher, and G. Sriram, 'Artificial Intelligence in Product Management: Systematic Review', *Int. Res. J. Mod. Eng. Technol. Sci.*, vol. 4, no. 7, pp. 2582–5208, Jul. 2022.

[17] R. Reddy Suram and B. Namatherdhala, 'Principle of Artificial Intelligence in Product Management', *Int. Res. J. Mod. Eng. Technol. Sci.*, no. 10, pp. 2582–5208, Oct. 2022, doi: 10.56726/IRJMETS30786.

[18] A. Mahendra, 'AI Product Strategy', in *AI Startup Strategy: A Blueprint to Building Successful Artificial Intelligence Products from Inception to Exit*, A. Mahendra, Ed., Berkeley, CA: Apress, 2023, pp. 137–167. doi: 10.1007/978-1-4842-9502-1_5.

[19] D. Dennehy, L. Kasraian, P. O'Raghallaigh, and K. Conboy, 'Product Market Fit Frameworks for Lean Product Development', presented at the R&D Management Conference 2016 "From Science to Society: Innovation and Value Creation", Cambridge, United Kingdom, 2016.

[20] I. M. Enholm, E. Papagiannidis, P. Mikalef, and J. Krogstie, 'Artificial Intelligence and Business Value: a Literature Review', *Inf. Syst. Front.*, vol. 24, no. 5, pp. 1709–1734, Oct. 2022, doi: 10.1007/s10796-021-10186-w.

[21] M. Trier *et al.*, 'Digital Responsibility: A Multilevel Framework for Responsible Digitalization', *Bus. Inf. Syst. Eng.*, vol. 65, no. 4, pp. 463–474, Aug. 2023, doi: 10.1007/s12599-023-00822-x.

[22] C. Rosenkranz, H. Vranešić, and R. Holten, 'Boundary Interactions and Motors of Change in Requirements Elicitation: A Dynamic Perspective on Knowledge Sharing', *J. Assoc. Inf. Syst.*, vol. 15, no. 6, Jun. 2014, doi: 10.17705/1jais.00364.

[23] R. R. Bond, M. Mulvenna, and H. Wang, 'Human Centered Artificial Intelligence: Weaving UX into Algorithmic Decision Making', in *Proceedings of RoCHI 2019*, Bucharest, 2019, pp. 2–9.

[24] M. O. Riedl, 'Human-centered artificial intelligence and machine learning', *Hum. Behav. Emerg. Technol.*, vol. 1, no. 1, pp. 33–36, 2019, doi: 10.1002/hbe2.117.

[25] W. Xu, 'Toward human-centered AI: a perspective from human-computer interaction', *Interactions*, vol. 26, no. 4, pp. 42–46, Jun. 2019, doi: 10.1145/3328485.

[26] B. Shneiderman, 'Human-Centered Artificial Intelligence: Reliable, Safe & Trustworthy', *Int. J. Human–Computer Interact.*, vol. 36, no. 6, pp. 495–504, Apr. 2020, doi: 10.1080/10447318.2020.1741118.

[27] B. Shneiderman, 'Responsible AI: bridging from ethics to practice', *Commun. ACM*, vol. 64, no. 8, pp. 32–35, Jul. 2021, doi: 10.1145/3445973.

[28] J. Auernhammer, 'Human-centered AI: The role of Human-centered Design Research in the development of AI', presented at the Design Research Society Conference 2020, Griffith University, Brisbane, Australia: Proceedings of DRS2020 International Conference, Vol. 1: Synergy, Situations, Aug. 2020. doi: 10.21606/drs.2020.282.

[29] A. Schmidt, 'Interactive Human Centered Artificial Intelligence: A Definition and Research Challenges', in *Proceedings of the International Conference on Advanced Visual Interfaces*, in AVI '20. New York, NY, USA: Association for Computing Machinery, Oct. 2020, pp. 1–4. doi: 10.1145/3399715.3400873.

[30] L. Fabri, B. Häckel, A. M. Oberländer, M. Rieg, and A. Stohr, 'Disentangling Human-AI Hybrids: Conceptualizing the Interworking of Humans and AI-Enabled Systems', *Bus. Inf. Syst. Eng.*, May 2023, doi: 10.1007/s12599-023-00810-1.

[31] UCL, 'Exploring Human Centred Explainable Artificial Intelligence', UCL Grand Challenges. Accessed: Feb. 07, 2024. [Online]. Available: https://www.ucl.ac.uk/grand-challenges/case-studies/2018/sep/exploring-human-centred-explainable-artificial-intelligence

[32] IBM, 'What is human-centered AI?', IBM Research Blog. Accessed: Feb. 07, 2024. [Online]. Available: https://research.ibm.com/blog/what-is-human-centered-ai

[33] University of Cambridge, 'Home - CHIA', Centre for Human-Inspired Artificial Intelligence. Accessed: Feb. 07, 2024. [Online]. Available: https://wp.chia.cam.ac.uk/

[34] Google Research, 'People + AI Research'. Accessed: Feb. 07, 2024. [Online]. Available: https://pair.withgoogle.com

[35] MIT Media Lab, 'Group Overview ‹ Media Lab Research Theme: Life with AI', MIT Media Lab. Accessed: Feb. 07, 2024. [Online]. Available: https://www.media.mit.edu/groups/media-lab-research-theme-life-with-ai/overview/

[36] Stanford, 'Home | Stanford HAI'. Accessed: Feb. 07, 2024. [Online]. Available: https://hai.stanford.edu/

[37] UC Berkeley, 'Center for Human-Compatible Artificial Intelligence – Center for Human-Compatible AI is building exceptional AI for humanity'. Accessed: Feb. 07, 2024. [Online]. Available: https://humancompatible.ai/

[38] University of Maryland, 'Human-Centered Artificial Intelligence', Human-Computer Interaction Lab. Accessed: Feb. 07, 2024. [Online]. Available: https://hcil.umd.edu/human-centered-ai/

[39] M. Schmitt, 'Automated machine learning: AI-driven decision making in business analytics', *Intell. Syst. Appl.*, vol. 18, p. 200188, May 2023, doi: 10.1016/j.iswa.2023.200188.

[40] X. He, K. Zhao, and X. Chu, 'AutoML: A Survey of the State-of-the-Art', *Knowl.-Based Syst.*, vol. 212, p. 106622, Jan. 2021, doi: 10.1016/j.knosys.2020.106622.

[41] D. Horneber and S. Laumer, 'Algorithmic Accountability', *Bus. Inf. Syst. Eng.*, vol. 66, no. 5, pp. 1805–1825, May 2023, doi: 10.1007/s12599-023-00817-8.

[42] A. Crisan and B. Fiore-Gartland, 'Fits and Starts: Enterprise Use of AutoML and the Role of Humans in the Loop', in *CHI Conference on Human Factors in Computing Systems (CHI '21)*, Yokohama, Japan, May 2021. Accessed: Oct. 07, 2023. [Online]. Available: http://arxiv.org/abs/2101.04296

[43] G. Margetis, S. Ntoa, M. Antona, and C. Stephanidis, 'Human-Centered Design of Artificial Intelligence', in *HANDBOOK OF HUMAN FACTORS AND ERGONOMICS*, John Wiley & Sons, Ltd, 2021, pp. 1085–1106. doi: 10.1002/9781119636113.ch42.

[44] D. Wang, X. Ma, and A. Y. Wang, 'Human-Centered AI for Data Science: A Systematic Approach'. arXiv, 2021. doi: 10.48550/ARXIV.2110.01108.

[45] S. Lins, K. D. Pandl, H. Teigeler, S. Thiebes, C. Bayer, and A. Sunyaev, 'Artificial Intelligence as a Service: Classification and Research Directions', *Bus. Inf. Syst. Eng.*, vol. 63, no. 4, pp. 441–456, Aug. 2021, doi: 10.1007/s12599-021-00708-w.

[46] V. K. Singh and K. Joshi, 'Automated Machine Learning (AutoML): an overview of opportunities for application and research', *J. Inf. Technol. Case Appl. Res.*, vol. 24, no. 2, pp. 75–85, Apr. 2022, doi: 10.1080/15228053.2022.2074585.

[47] A. Burgess, 'AI Prototyping', in *The Executive Guide to Artificial Intelligence: How to identify and implement applications for AI in your organization*, A. Burgess, Ed., Cham: Springer International Publishing, 2018, pp. 117–127. doi: 10.1007/978-3-319-63820-1_7.

[48] Y. Harviv and N. Gift, *Implementing MLOps in the Enterprise: A Production-First Approach*, 1st ed. O'Reilly Media, 2023.

[49] G. Smith, M. Papadopoulos, J. Sanz, M. Grech, and H. Norris, 'Unleashing innovation using low code/no code – The age of the citizen developer'. 2020. Accessed: Nov. 06, 2023. [Online]. Available: https://www.adlittle-lebanon.com/sites/default/files/prism/arthur_d_little_prism_low_code_no_code.pdf

[50] M. Gillies *et al.*, 'Human-Centred Machine Learning', in *Proceedings of the 2016 CHI Conference Extended Abstracts on Human Factors in Computing Systems*, San Jose California USA: ACM, May 2016, pp. 3558–3565. doi: 10.1145/2851581.2856492.

[51] G. Dove, K. Halskov, J. Forlizzi, and J. Zimmerman, 'UX Design Innovation: Challenges for Working with Machine Learning as a Design Material', in *Proceedings of the 2017 CHI Conference on Human Factors in Computing Systems*, Denver Colorado USA: ACM, May 2017, pp. 278–288. doi: 10.1145/3025453.3025739.

[52] P. van Allen, 'Prototyping ways of prototyping AI', *Interactions*, vol. 25, no. 6, pp. 46–51, Oct. 2018, doi: 10.1145/3274566.

[53] Q. Yang, 'Profiling Artificial Intelligence as a Material for User Experience Design', in *CHI EA '21: Extended Abstracts of the 2021 CHI Conference on Human Factors in Computing Systems*, Pittsburgh PA USA: Carnegie Mellon University, 2020. doi: https://doi.org/10.1145/3411763.3457783.

[54] M. Glintschert, *AI-driven IT and its Potentials - a State-of-the-Art Approach*. 2020. doi: 10.13140/RG.2.2.34726.68163.

[55] H. Subramonyam, C. Seifert, and E. Adar, 'ProtoAI: Model-Informed Prototyping for AI-Powered Interfaces', in *26th International Conference on Intelligent User Interfaces*, in IUI '21. New York, NY, USA: Association for Computing Machinery, Apr. 2021, pp. 48–58. doi: 10.1145/3397481.3450640.

[56] E. Jiang *et al.*, 'PromptMaker: Prompt-based Prototyping with Large Language Models', in *CHI Conference on Human Factors in Computing Systems Extended Abstracts*, New Orleans LA USA: ACM, Apr. 2022, pp. 1–8. doi: 10.1145/3491101.3503564.

[57] S. Moore, Q. V. Liao, and H. Subramonyam, 'fAIlureNotes: Supporting Designers in Understanding the Limits of AI Models for Computer Vision Tasks', in *CHI '23: Proceedings of the 2023 CHI Conference on Human Factors in Computing System*, Hamburg, 28.4 2023, pp. 1–19. Accessed: Nov. 05, 2023. [Online]. Available: https://dl.acm.org/doi/10.1145/3544548.3581242

[58] V. Dibia, A. Cox, and J. Weisz, 'Designing for Democratization: Introducing Novices to Artificial Intelligence Via Maker Kits'. arXiv, Jan. 05, 2019. doi: https://doi.org/10.48550/arXiv.1805.10723.

[59] N. Malsattar, T. Kihara, and E. Giaccardi, 'Designing and Prototyping from the Perspective of AI in the Wild', in *Proceedings of the 2019 on Designing Interactive Systems Conference*, in DIS '19. New York, NY, USA: Association for Computing Machinery, Jun. 2019, pp. 1083–1088. doi: 10.1145/3322276.3322351.

[60] H. Cramer and J. Kim, 'Confronting the tensions where UX meets AI', *Interactions*, vol. 26, no. 6, pp. 69–71, Oct. 2019, doi: 10.1145/3364625.

[61] T. Posner and L. Fei-Fei, 'AI will change the world, so it's time to change AI', *Nature*, vol. 588, no. 7837, pp. S118–S118, Dec. 2020, doi: 10.1038/d41586-020-03412-z.

[62] N. B. Yams and G. E. Shubina, 'What AI Can Do for Innovation Managers and Innovation Managers for AI', in *Series on Technology Management*, vol. 38, WORLD SCIENTIFIC (EUROPE), 2022, pp. 11–36. doi: 10.1142/9781800611337_0002.

[63] H. Subramonyam, J. Im, C. Seifert, and E. Adar, 'Solving Separation-of-Concerns Problems in Collaborative Design of Human-AI Systems through Leaky Abstractions', in *Proceedings of the 2022 CHI Conference on Human Factors in Computing Systems*, in CHI '22. New York, NY, USA: Association for Computing Machinery, Apr. 2022, pp. 1–21. doi: 10.1145/3491102.3517537.

[64] I. Alves, L. A. F. Leite, P. Meirelles, F. Kon, and C. S. R. Aguiar, 'Practices for Managing Machine Learning Products: A Multivocal Literature Review', *IEEE Trans. Eng. Manag.*, pp. 1–31, 2023, doi: 10.1109/TEM.2023.3287759.

[65] K. J. K. Feng, M. J. Coppock, and D. W. McDonald, 'How Do UX Practitioners Communicate AI as a Design Material? Artifacts, Conceptions, and Propositions', in *Proceedings of the 2023 ACM Designing Interactive Systems Conference*, Pittsburgh PA USA: ACM, Jul. 2023, pp. 2263–2280. doi: 10.1145/3563657.3596101.

[66] L. Sundberg and J. Holmström, 'Democratizing artificial intelligence: How no-code AI can leverage machine learning operations', *Bus. Horiz.*, vol. 66, no. 6, pp. 777–788, Nov. 2023, doi: 10.1016/j.bushor.2023.04.003.

[67] M. Hartikainen, K. Väänänen, A. Lehtiö, S. Ala-Luopa, and T. Olsson, 'Human-Centered AI Design in Reality: A Study of Developer Companies' Practices: A study of Developer Companies' Practices', in *Nordic Human-Computer Interaction Conference*, Aarhus Denmark: ACM, Oct. 2022, pp. 1–11. doi: 10.1145/3546155.3546677.

[68] T. Wu *et al.*, 'PromptChainer: Chaining Large Language Model Prompts through Visual Programming', in *Extended Abstracts of the 2022 CHI Conference on Human Factors in Computing Systems*, in CHI EA '22. New York, NY, USA: Association for Computing Machinery, Apr. 2022, pp. 1–10. doi: 10.1145/3491101.3519729.

[69] S. Petridis, M. Terry, and C. J. Cai, 'PromptInfuser: Bringing User Interface Mock-ups to Life with Large Language Models', in *Extended Abstracts of the 2023 CHI Conference on Human Factors in Computing Systems*, in CHI EA '23. New York, NY, USA: Association for Computing Machinery, Apr. 2023, pp. 1–6. doi: 10.1145/3544549.3585628.

[70] J. T. Browne, 'Wizard of Oz Prototyping for Machine Learning Experiences', in *Extended Abstracts of the 2019 CHI Conference on Human Factors in Computing Systems*, in CHI EA '19. New York, NY, USA: Association for Computing Machinery, May 2019, pp. 1–6. doi: 10.1145/3290607.3312877.

[71] H. Scurto, B. Caramiaux, and F. Bevilacqua, 'Prototyping Machine Learning Through Diffractive Art Practice', in *Designing Interactive Systems Conference 2021*, Virtual Event USA: ACM, Jun. 2021, pp. 2013–2025. doi: 10.1145/3461778.3462163.

[72] A. Esposito *et al.*, 'End-User Development for Artificial Intelligence: A Systematic Literature Review', in *End-User Development*, L. D. Spano, A. Schmidt, C. Santoro, and S. Stumpf, Eds., in Lecture Notes in Computer Science. Cham: Springer Nature Switzerland, 2023, pp. 19–34. doi: 10.1007/978-3-031-34433-6_2.

[73] R. Stackowiak and T. Kelly, *Design Thinking in Software and AI Projects: Proving Ideas Through Rapid Prototyping*. Berkeley, CA: Apress, 2020. doi: 10.1007/978-1-4842-6153-8.

[74] S. Nalchigar and E. Yu, 'Designing Business Analytics Solutions: A Model-Driven Approach', *Bus. Inf. Syst. Eng.*, vol. 62, no. 1, pp. 61–75, Feb. 2020, doi: 10.1007/s12599-018-0555-z.

[75] M. K. Hong, A. Fourney, D. DeBellis, and S. Amershi, 'Planning for Natural Language Failures with the AI Playbook', in *Proceedings of the 2021 CHI Conference on Human Factors in Computing Systems*, Yokohama Japan: ACM, May 2021, pp. 1–11. doi: 10.1145/3411764.3445735.

[76] A. Rizzo, F. Montefoschi, M. Caporali, A. Gisondi, G. Burresi, and R. Giorgi, 'Rapid prototyping IoT solutions based on Machine Learning', in *Proceedings of the European Conference on Cognitive Ergonomics 2017*, Umeå Sweden: ACM, Sep. 2017, pp. 184–187. doi: 10.1145/3121283.3121291.

[77] F. Bernardo, M. Zbyszyński, M. Grierson, and R. Fiebrink, 'Designing and Evaluating the Usability of a Machine Learning API for Rapid Prototyping Music Technology', *Front. Artif. Intell.*, vol. 3, 2020, doi: https://doi.org/10.3389/frai.2020.00013.

[78] V. Bilgram and F. Laarmann, 'Accelerating Innovation With Generative AI: AI-Augmented Digital Prototyping and Innovation Methods', *IEEE Eng. Manag. Rev.*, vol. 51, no. 2, pp. 18–25, 2023, doi: 10.1109/EMR.2023.3272799.

[79] T. Nagarajah and G. Poravi, 'A Review on Automated Machine Learning (AutoML) Systems', in *2019 IEEE 5th International Conference for Convergence in Technology (I2CT)*, Mar. 2019, pp. 1–6. doi: 10.1109/I2CT45611.2019.9033810.

[80]  J. Bender and J. Ovtcharova, 'Prototyping Machine-Learning-Supported Lead Time Prediction Using AutoML', *Procedia Comput. Sci.*, vol. 180, pp. 649–655, 2021, doi: 10.1016/j.procs.2021.01.287.

[81]  F. Calefato, L. Quaranta, F. Lanubile, and M. Kalinowski, 'Assessing the Use of AutoML for Data-Driven Software Engineering', in *Proceeding of 17th ACM/IEEE International Symposium on Empirical Software Engineering and Measurement*, Abu Dhabi, Dec. 2023. Accessed: Oct. 07, 2023. [Online]. Available: http://arxiv.org/abs/2307.10774

[82]  M. Truss and S. Böhm, 'AI-based Classification of Customer Support Tickets: State of the Art and Implementation with AutoML', in *Proceedings of the IWEMB 2021 and 2022*, Leipzig: Publiqation, 2023.

[83]  C. Brandon and T. Margaria, 'Low-Code/No-Code Artificial Intelligence Platforms for the Health Informatics Domain', in *Electronic Communications of the EASST*, Berlin, 2022.

[84]  L. Sun, Z. Zhou, W. Wu, Y. Zhang, R. Zhang, and W. Xiang, 'Developing a Toolkit for Prototyping Machine Learning-Empowered Products':, *Int. J. Des.*, vol. 14, no. 2, 2020.

[85]  C. T. Wolf, 'Democratizing AI? experience and accessibility in the age of artificial intelligence', *XRDS Crossroads ACM Mag. Stud.*, vol. 26, no. 4, pp. 12–15, Jul. 2020, doi: 10.1145/3398370.

[86]  C. V. K. Iyer *et al.*, 'Trinity: A No-Code AI platform for complex spatial datasets', in *Proceedings of the 4th ACM SIGSPATIAL International Workshop on AI for Geographic Knowledge Discovery*, in GEOAI '21. New York, NY, USA: Association for Computing Machinery, Nov. 2021, pp. 33–42. doi: 10.1145/3486635.3491072.

[87]  R. Du *et al.*, 'Experiencing Visual Blocks for ML: Visual Prototyping of AI Pipelines'. 2023. Accessed: Nov. 06, 2023. [Online]. Available: https://research.google/pubs/pub52583/

[88]  N. Pfeuffer *et al.*, 'Explanatory Interactive Machine Learning: Establishing an Action Design Research Process for Machine Learning Projects', *Bus. Inf. Syst. Eng.*, Apr. 2023, doi: 10.1007/s12599-023-00806-x.

[89]  D. Rall, B. Bauer, and T. Fraunholz, 'Towards Democratizing AI: A Comparative Analysis of AI as a Service Platforms and the Open Space for Machine Learning Approach', in *Proceedings of the 2023 7th International Conference on Cloud and Big Data Computing*, in ICCBDC '23. New York, NY, USA: Association for Computing Machinery, Oct. 2023, pp. 34–39. doi: 10.1145/3616131.3616136.

[90]  K. Peffers, T. Tuunanen, M. A. Rothenberger, and S. Chatterjee, 'A Design Science Research Methodology for Information Systems Research', *J. Manag. Inf. Syst.*, vol. 24, no. 3, pp. 45–77, Dec. 2007, doi: 10.2753/MIS0742-1222240302.

[91]  A. Cleven, P. Gubler, and K. M. Hüner, 'Design alternatives for the evaluation of design science research artifacts', in *Proceedings of the 4th International Conference on Design Science Research in Information Systems and Technology - DESRIST '09*, Philadelphia, Pennsylvania: ACM Press, 2009, p. 1. doi: 10.1145/1555619.1555645.

[92]  K. Peffers, M. Rothenberger, T. Tuunanen, and R. Vaezi, 'Design Science Research Evaluation', in *Design Science Research in Information Systems. Advances in Theory and Practice*, vol. 7286, K. Peffers, M. Rothenberger, and B. Kuechler, Eds., in Lecture Notes in Computer Science, vol. 7286. , Berlin, Heidelberg: Springer Berlin Heidelberg, 2012, pp. 398–410. doi: 10.1007/978-3-642-29863-9_29.

[93]  J. Venable, J. Pries-Heje, and R. Baskerville, 'FEDS: a Framework for Evaluation in Design Science Research', *Eur. J. Inf. Syst.*, vol. 25, no. 1, pp. 77–89, Jan. 2016, doi: 10.1057/ejis.2014.36.

[94]  G. Berg, 'Image Classification with Machine Learning as a Service', Bachelor Thesis, Linnaeus University, Växjö, SWE, 2022. [Online]. Available: https://www.diva-portal.org/smash/get/diva2:1667596/FULLTEXT01.pdf

[95]  W. Choi, T. Choi, and S. Heo, 'A Comparative Study of Automated Machine Learning Platforms for Exercise Anthropometry-Based Typology Analysis: Performance Evaluation of AWS SageMaker, GCP VertexAI, and MS Azure', *Bioengineering*, vol. 10, no. 8, p. 891, Jul. 2023, doi: 10.3390/bioengineering10080891.

[96]  A. Nisanova *et al.*, 'Automated machine learning to predict anatomical outcomes in pneumatic retinopexy', *Invest. Ophthalmol. Vis. Sci.*, vol. 64, no. 8, p. 223, Jun. 2023.

[97]  Y. Sun and P. B. Kantor, 'Cross-Evaluation: A new model for information system evaluation', *J. Am. Soc. Inf. Sci. Technol.*, vol. 57, no. 5, pp. 614–628, 2006, doi: 10.1002/asi.20324.

[98]  E. Gummesson, 'Qualitative research in management: addressing complexity, context and persona', *Manag. Decis.*, vol. 44, no. 2, pp. 167–179, Jan. 2006, doi: 10.1108/00251740610650175.

[99]  J. Venable, J. Pries-Heje, and R. Baskerville, 'A Comprehensive Framework for Evaluation in Design Science Research', in *Design Science Research in Information Systems. Advances in Theory and Practice*, vol. 7286, K. Peffers, M. Rothenberger, and B. Kuechler, Eds., in Lecture Notes in Computer Science, vol. 7286. , Berlin, Heidelberg: Springer Berlin Heidelberg, 2012, pp. 423–438. doi: 10.1007/978-3-642-29863-9_31.

[100] J. F. Nunamaker, M. Chen, and T. D. M. Purdin, 'Systems Development in Information Systems Research', *J. Manag. Inf. Syst.*, vol. 7, no. 3, pp. 89–106, Dec. 1990, doi: 10.1080/07421222.1990.11517898.

[101] Hevner, March, Park, and Ram, 'Design Science in Information Systems Research', *MIS Q.*, vol. 28, no. 1, p. 75, 2004, doi: 10.2307/25148625.

[102] J. Iivari, 'A Paradigmatic Analysis of Information Systems As a Design Science', *Scand. J. Inf. Syst.*, vol. 19, no. 2, Jan. 2007, [Online]. Available: https://aisel.aisnet.org/sjis/vol19/iss2/5

[103] Google, 'Documentation of Vertex AI', Documentation of Vertex AI. Accessed: Jan. 01, 2024. [Online]. Available: https://cloud.google.com/vertex-ai/docs?hl=en

[104] Amazon Web Services, 'No-code Machine Learning - Amazon SageMaker Canvas - AWS', Amazon Web Services, Inc. Accessed: Jan. 01, 2024. [Online]. Available: https://aws.amazon.com/sagemaker/canvas/

[105] Microsoft, 'Azure Automated Machine Learning - AutoML | Microsoft Azure', Azure Automated Machine Learning - AutoML | Microsoft Azure. Accessed: Jan. 01, 2024. [Online]. Available: https://azure.microsoft.com/en-us/products/machine-learning/automatedml/

[106] G. Ramos, C. Meek, P. Simard, J. Suh, and S. Ghorashi, 'Interactive machine teaching: a human-centered approach to building machine-learned models', *Human–Computer Interact.*, vol. 35, no. 5–6, pp. 413–451, Nov. 2020, doi: 10.1080/07370024.2020.1734931.

[107] A. Pentland, A. Lipton, and T. Hardjono, *Building the new economy: data as capital*. in MIT connection science & engineering. Cambridge, Massachusetts London, England: The MIT Press, 2021.

[108] A. Barbala, T. Sporsem, and V. Stray, 'Data-Driven Development in Public Sector: How Agile Product Teams Maneuver Data Privacy Regulations', in *Agile Processes in Software Engineering and Extreme Programming*, C. J. Stettina, J. Garbajosa, and P. Kruchten, Eds., in Lecture Notes in Business Information Processing. Cham: Springer Nature Switzerland, 2023, pp. 165–180. doi: 10.1007/978-3-031-33976-9_11.

[109] A. Schmid and M. Wiesche, 'The importance of an ethical framework for trust calibration in AI', *IEEE Intell. Syst.*, pp. 1–8, 2023, doi: 10.1109/MIS.2023.3320443.

[110] A. Schmid and M. Wiesche, 'Building Trust in AI: A new Understanding of the Role of Reliability', *AMCIS 2023 Proc.*, Aug. 2023, [Online]. Available: https://aisel.aisnet.org/amcis2023/sig_aiaa/sig_aiaa/12

[111] T. Capel and M. Brereton, 'What is Human-Centered about Human-Centered AI? A Map of the Research Landscape', in *Proceedings of the 2023 CHI Conference on Human Factors in Computing Systems*, Hamburg Germany: ACM, Apr. 2023, pp. 1–23. doi: 10.1145/3544548.3580959.

[112] D. Cotroneo, C. Improta, P. Liguori, and R. Natella, 'Vulnerabilities in AI Code Generators: Exploring Targeted Data Poisoning Attacks'. arXiv, Nov. 06, 2023. doi: https://doi.org/10.48550/arXiv.2308.04451.